\documentclass[jkps,preprint,fleqn,showpacs,showkeys]{revtex4}
\usepackage{graphicx}
\usepackage{amssymb}
\usepackage{amsmath}
\usepackage{bm}
\begin{document}
\setcounter{page}{0}
\title{Generation of a bubble universe and the information loss problem}
\author{Dong-han \surname{Yeom}\footnote{A proceeding for IKSRA09. Talk on November 2, 2009, Seoul, Korea.}}
\email{innocent@muon.kaist.ac.kr}
\thanks{Fax: +82-42-350-5537}
\affiliation{Department of Physics, KAIST, Daejeon 305-701, Republic of Korea}

\date[]{Received \today}

\begin{abstract}
We discuss a possible scenario to induce a bubble universe using buildable bubbles. First, we observe that if there is no violation of the null energy condition, all bubbles are buildable and do not have inflation if their initial conditions are assigned along the in-going null surface in an almost flat background. However, if the null energy condition is sufficiently violated, then the generation of a bubble universe is possible. Second, the author will observe that a sufficient violation of the null energy condition may be possible using a buildable bubble due to the Hawking radiation of a de Sitter space. Note that buildable bubbles seem surely to be allowed in quantum gravity. Therefore, if this scenario is correct, the generation of a bubble universe and the violation of unitarity should be allowed in quantum gravity.
\end{abstract}


\pacs{04.62.+v, 04.70.-s, 04.20.-q, 04.25.Dm}

\keywords{false vacuum bubbles, bubble universes, information loss}

\maketitle

\section{INTRODUCTION}

One of the most important problems of quantum gravity is the information loss problem in black holes \cite{Hawking:1976ra}. If there is information loss, it will imply that the time evolution of a black hole is no longer unitary, and hence quantum gravity may not be unitary.

This problem is more serious in a bubble universe. If various false vacua and tunneling between each vacuum is allowed \cite{Susskind:2003kw}, one has to consider a false vacuum bubble in a true vacuum background \cite{Blau:1986cw}\cite{Sato:1981bf}\cite{Aguirre:2005xs}\cite{Alberghi:1999kd}\cite{Lee:2006vka}. If the bubble is greater than the true vacuum background, there is no problem of information loss. However, if the bubble is smaller than the true vacuum background but the bubble contains inflation, there will be a bubble universe that is separated from the outer asymptotic infinity. Between two asymptotic infinities, there will be a wormhole (e.g., a Schwarzschild wormhole) and the wormhole will form a black hole. As the black hole evaporates, two asymptotic regions will be separated forever.

If quantum gravity is a unitary theory, one possible interpretation is that the generation of a bubble universe is allowed but there is no information loss for all observers. If one trusts observer complementarity \cite{Susskind:1993if}, this may be helpful for the information loss problem. However, some authors have observed that the generation of a bubble universe contradicts holography or black hole complementarity \cite{Freivogel:2005qh}. Also, there are some studies in which it is asserted that observer complementarity cannot be a fundamental resolution of the information loss problem \cite{Yeom:2008qw}\cite{Ge:2005bn}.

Then, the next possible interpretation is that a bubble universe should not be generated, and hence the initial conditions that induce a bubble universe should not be allowed. One important observation is that, whenever one assumes global hyperbolicity and the null energy condition, if a bubble contains inflation, the bubble should begin from a singular state in terms of general relativity, and hence be unbuildable in a general relativistic sense \cite{Farhi:1986ty}. Therefore, to induce such initial conditions, some authors have considered quantum tunneling \cite{Farhi:1989yr}. The authors could calculate the probability using a reasonable approximation of quantum gravity. However, still other authors doubt the possibility of tunneling for an unbuildable bubble \cite{Freivogel:2005qh}.

In this paper, the author discusses a possible scenario to induce a bubble universe using \textit{buildable} bubbles. First, the author will observe that if there is no violation of the null energy condition, all bubbles are buildable and do not have inflation if their initial conditions are assigned along the in-going null surface in an almost flat background. However, if the null energy condition is sufficiently violated, then the generation of a bubble universe is possible. Second, the author will observe that a sufficient violation of the null energy condition may be possible using a buildable bubble due to the Hawking radiation of a de Sitter space \cite{Gibbons:1977mu}. Note that buildable bubbles seem surely to be allowed in quantum gravity. Therefore, if this scenario is correct, the generation of a bubble universe and the violation of unitarity should be allowed in quantum gravity.

\section{Dynamics of false vacuum bubbles}

\begin{figure}
\begin{center}
\includegraphics[scale=0.45]{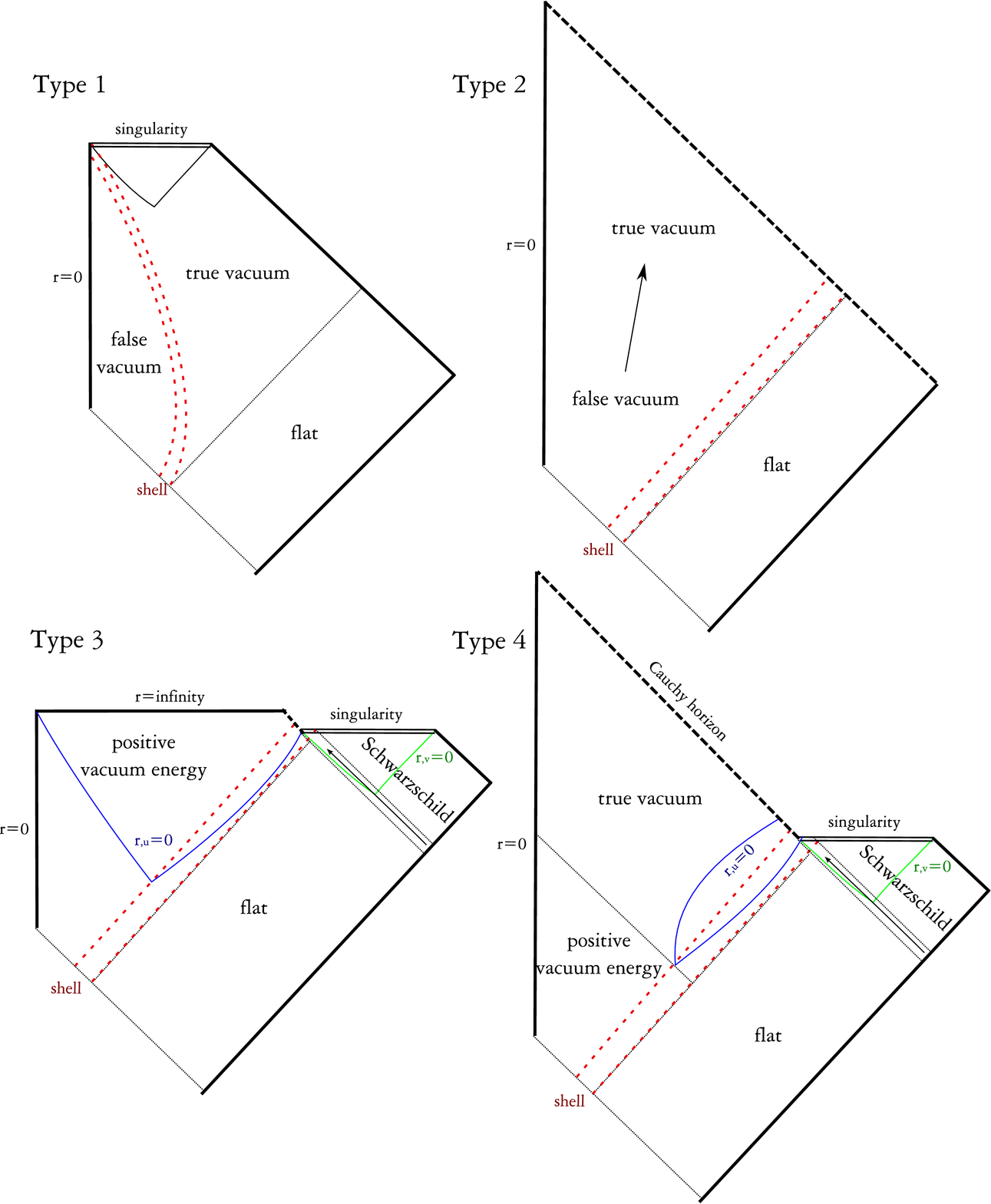}
\caption{\label{fig:sol}Dynamics of false vacuum bubbles.}
\end{center}
\end{figure}

The dynamics of false vacuum bubbles were found using the thin shell approximation \cite{Israel:1966rt}\cite{Blau:1986cw}. The thin shell approximation assumes that the inside of a bubble has a de Sitter metric and the outside of a bubble has a Schwarzschild metric. Between the two regions, there is a thin mass shell. The shell can collapse or expand. If it expands, the inside of the bubble includes inflation, but the initial state is unbuildable \cite{Farhi:1986ty}. Therefore, some authors have considered tunneling between a collapsing bubble and an expanding bubble \cite{Farhi:1989yr}.

In one paper \cite{Hansen:2009kn}, the author and colleagues studied the dynamics of false vacuum bubbles beyond the thin shell approximation using numerical simulations; the technique had been previously used by many authors \cite{Piran}\cite{Hong:2008mw}. We assumed a massless scalar field $\Phi$ and a potential $V(\Phi)$. We assume that $\Phi=0$ is a true vacuum and $\Phi=\Phi_{0}$ has non-zero vacuum energy. Also, we assume the spherical symmetric metric:
\begin{eqnarray}
ds^{2} = -\alpha^{2}(u,v) du dv + r^{2}(u,v) d\Omega^{2},
\end{eqnarray}
where $u$ increases along the in-going null direction and $v$ increases along the out-going null direction. Along the initial in-going null surface $v=v_{\mathrm{i}}$, we assumed the initial false vacuum bubble using the field configuration:
\begin{eqnarray}
\Phi(u,v_{\mathrm{i}}) = \left\{ \begin{array}{ll}
0 & u < u_{\mathrm{shell}},\\
\Phi_{0} G(u) & u_{\mathrm{shell}} \leq u < u_{\mathrm{shell}}+\Delta u,\\
\Phi_{0} & u_{\mathrm{shell}}+\Delta u \leq u,
\end{array} \right.
\end{eqnarray}
where $G(u)$ is a pasting function that goes from $0$ to $1$ by a smooth way. We choose $G(u)$ by
\begin{eqnarray}
G(u) = \sin^{2} \left[\frac{\pi(u-u_{\mathrm{shell}})}{2\Delta u}\right].
\end{eqnarray}
If $u$ is greater than $u_{\mathrm{shell}}+\Delta u$, the region has non-zero vacuum energy. If $u$ is less than $u_{\mathrm{shell}}$, the region has zero vacuum energy. Between the two regions, there is a transition region or a massive shell and the thickness of the shell is initially $\Delta u$.

Figure~\ref{fig:sol} shows possible solutions. Type~1 in Figure~\ref{fig:sol} is a collapsing shell. As the energy of a shell increases, the shell tends to expand (Type~2). However, via the dynamics of the inside region, there was no inflation. This is obvious since
\begin{eqnarray}
r_{,uu}|_{r_{,u}=0} = - rW^{2} \leq 0,
\end{eqnarray}
and hence, the in-going observer cannot see the $r_{,u}=0$, if there is no violation of the null energy condition. If we violate the null energy condition using exotic matter fields, we could obtain the inflating solutions (Type~3 and Type~4). Note that if there is no violation of the null energy condition, all of initial field configurations are buildable. Therefore, if there is a sufficient combination of negative energy, a classically buildable field configuration can inflate.

\section{Generation of a negative energy bath}

To obtain a sufficient amount of negative energy, we may use Hawking radiation from a false vacuum bubble \cite{Gibbons:1977mu}. In $1+1$ dimensions, it can be easily calculated using the renormalized stress tensor \cite{Davies:1976ei}\cite{Birrell:1982ix}:
\begin{eqnarray} \label{semiclassical}
\langle T_{uu} \rangle &=& \frac{P}{\alpha^{2}}\left(\alpha \alpha_{,uu} - 2 {\alpha_{,u}}^{2}\right),\\
\langle T_{uv} \rangle = \langle T_{vu} \rangle &=& -\frac{P}{\alpha^{2}}\left(\alpha\alpha_{,uv}-\alpha_{,u}\alpha_{,v}\right),\\
\langle T_{vv} \rangle &=& \frac{P}{\alpha^{2}}\left(\alpha \alpha_{,vv} - 2 {\alpha_{,v}}^{2}\right),
\end{eqnarray}
where $P$ is a constant that is proportional to the number of independent modes of Hawking radiation. For a de Sitter space, $\langle T_{uu} \rangle = \langle T_{vv} \rangle \simeq -\Lambda$ \cite{Birrell:1982ix}. If we assume the S-wave approximation, we will obtain $\langle T_{uu} \rangle = \langle T_{vv} \rangle \simeq -\Lambda/r^{2}$ in spherically symmetric $3+1$ dimensions.

\begin{figure}
\begin{center}
\includegraphics[scale=0.45]{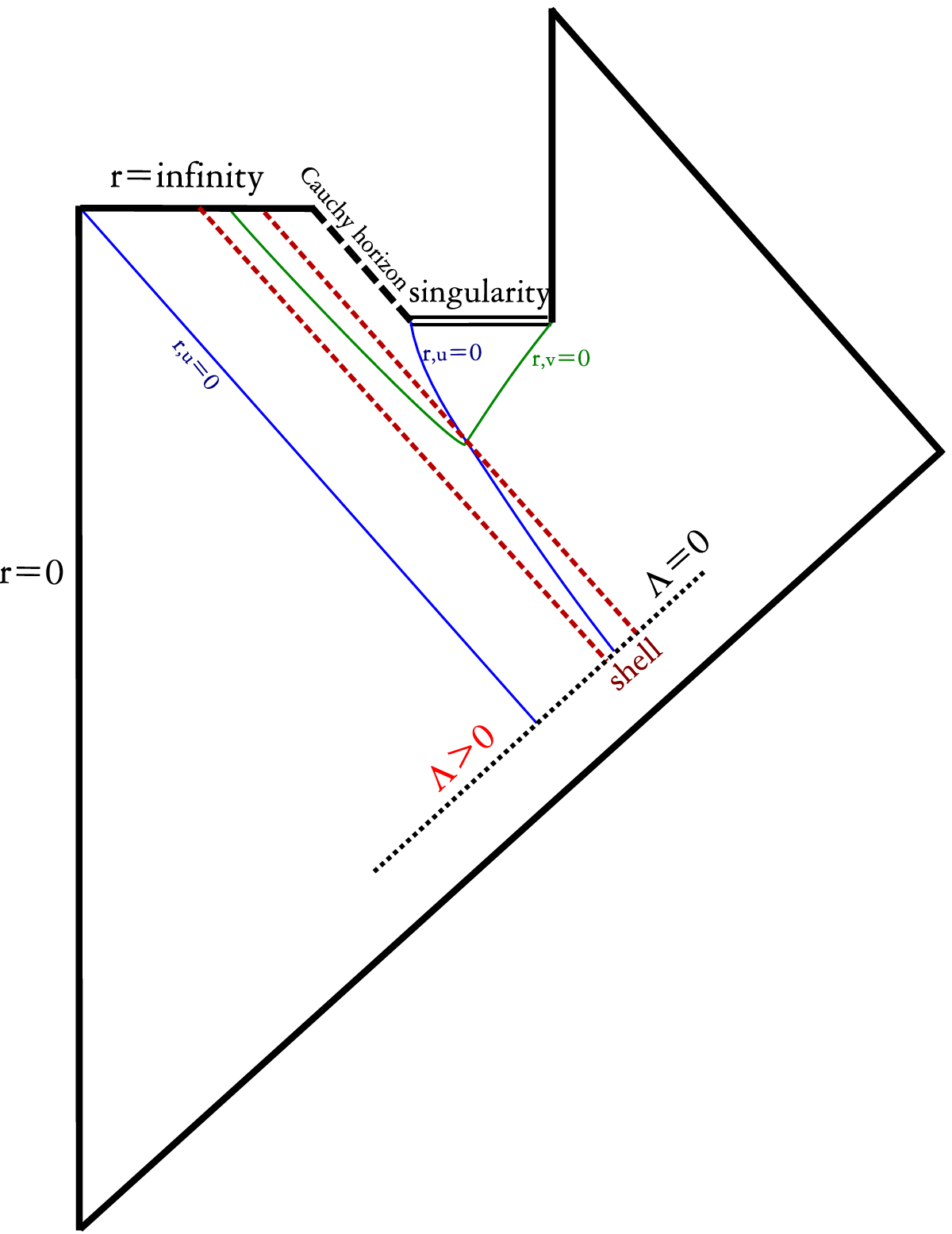}
\includegraphics[scale=0.45]{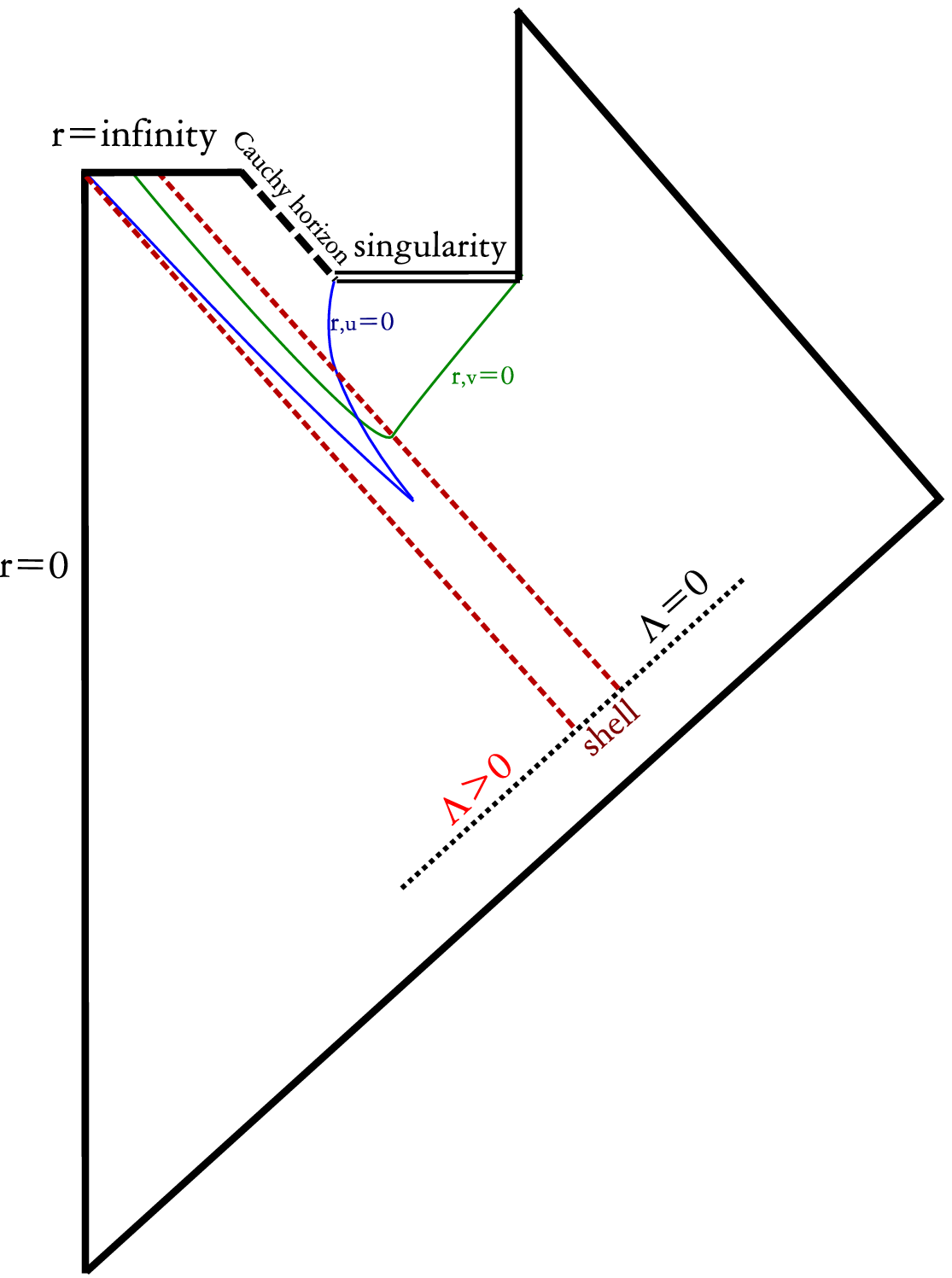}
\caption{\label{fig:unbuildable}Unbuildable bubbles.}
\end{center}
\end{figure}

\begin{figure}
\begin{center}
\includegraphics[scale=0.45]{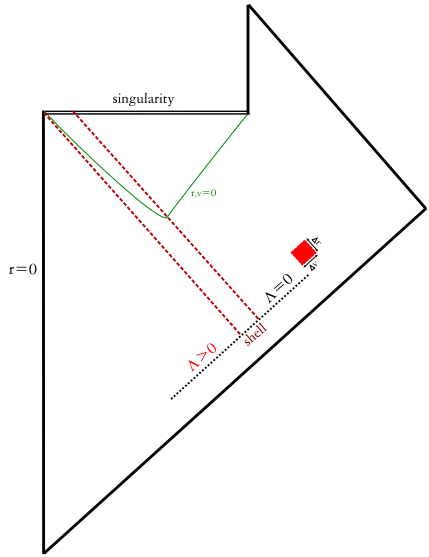}
\caption{\label{fig:buildable}Buildable bubbles. At the outside of the shell, during a short time, we can find a negative energy bath.}
\end{center}
\end{figure}

We guess that if we prepare a sufficiently large bubble along the initial out-going null surface, we may obtain the cosmological horizon and the violation of the null energy condition. However, if we impose the initial condition along the initial out-going null surface using tunneling, we may obtain unbuildable bubble solutions (Figure~\ref{fig:unbuildable}). However, if the shell is sufficiently smaller than the size of the cosmological horizon, one can obtain the buildable bubble solution (Figure~\ref{fig:buildable}). If there is a violation of the null energy condition outside of the shell in Figure~\ref{fig:unbuildable}, although there is no cosmological horizon in Figure~\ref{fig:buildable}, one can find the same order of violation in Figure~\ref{fig:buildable}, since we can smoothly deform from Figure~\ref{fig:unbuildable} to Figure~\ref{fig:buildable} by changing initial parameters \footnote{Figure~\ref{fig:unbuildable} and Figure~\ref{fig:buildable} will be numerically confirmed in the author's future work. However, these figures are still reasonable in terms of the thin shell approximation.}.

\begin{figure}
\begin{center}
\includegraphics[scale=0.45]{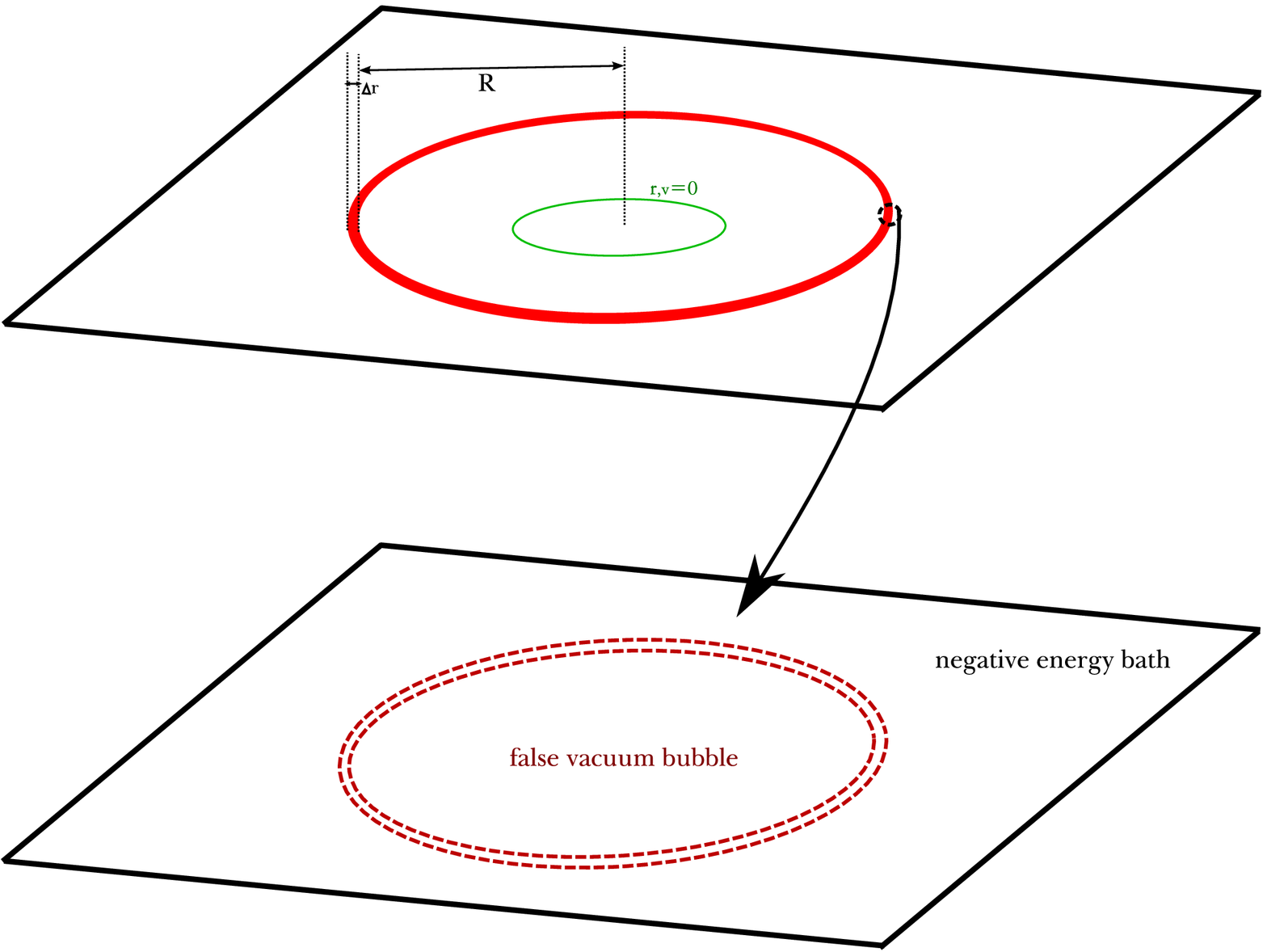}
\caption{\label{fig:bath}Using a negative energy bath.}
\end{center}
\end{figure}

Then, for a sufficiently large $R > r_{\mathrm{black~hole}}$, and for a sufficiently local space $\Delta r \ll R$ and a short time $\Delta v \ll R$, we can observe a negative energy bath (Figure~\ref{fig:buildable} and Figure~\ref{fig:bath}). Using the negative energy bath, we can induce a bubble universe. Of course, the question is whether the negative energy bath gives a sufficient amount of negative energy. The question is related to whether $\Lambda$ can be chosen arbitrarily. If the size of the shell is slightly smaller than the size of the cosmological horizon, the field combination for the shell will be buildable, though the vacuum energy is arbitrarily large. Note that although the vacuum energy is large as a simulation parameter, if we tune the unit length and the unit mass via a large number of massless fields, we can make our simulation semi-classical \cite{Yeom:2009zp}. Therefore, the size of the shell can be sufficiently larger than the Planck length, and hence it can be trustable in the semi-classical sense.

\section{CONCLUSIONS}

We need further calculations to confirm this scenario. We need to check whether all of our parameters are consistent and possible in the semi-classical regime. However, if the scenario is possible, it will imply that the generation of a bubble universe and a violation of unitarity are allowed in semi-classical physics. Here, we did not assume a special metric that already violated unitarity or exotic matters that were inserted by hand. Also, we assumed tunneling, which seemed surely to be allowed in quantum gravity. This scenario clearly simplifies the sufficient conditions to derive a bubble universe. If quantum gravity is unitary, one of our assumptions should be wrong. Hence, the confirmation of our scenario and the clarification of sufficient conditions for the generation of a bubble universe will be an interesting future work.

\begin{acknowledgments}
The author would like to thank Ewan Stewart, Dong-il Hwang, and Jakob Hansen for helpful discussions.
This work was supported by Korea Research Foundation grants (KRF-313-2007-C00164, KRF-341-2007-C00010) funded by the Korean government (MOEHRD) and BK21.
\end{acknowledgments}

\end{document}